\documentclass[12pt]{article}
\usepackage{epsfig}
\newcommand\bea{\begin{eqnarray}}
\newcommand\eea{\end{eqnarray}}
\begin{document}
\thispagestyle{empty}
\bibliographystyle{unsrt}
\setlength{\baselineskip}{18pt}
\parindent 24pt
\vspace{40pt}

\newcommand{\g}{\gamma}
\newcommand{\s}{\sigma}
\newcommand{\la}{\lambda}
\newcommand{\m}{\mu}
\newcommand{\om}{\omega}
\newcommand{\Om}{\Omega}
\newcommand{\de}{\delta}
\newcommand{\co}{\coth{\hbar\omega\over 2kT}}
\newcommand{\Co}{\coth^{2}{\hbar\omega\over 2kT}}

\begin{center}{
{\Large {{\bf Uncertainty functions of the open quantum harmonic
oscillator in the Lindblad theory }} }
\vskip 1truecm
A. Isar ${\dagger}{\ddagger}^{(a)}$ and W. Scheid $\ddagger$  \\
$\dagger$ {\it Department of Theoretical Physics, Institute of
Atomic Physics, \\
Bucharest-Magurele, Romania }\\
$\ddagger$ {\it Institut f\"{u}r Theoretische Physik der
Justus-Liebig-Universit\"{a}t,\\
Giessen, Germany} }
\end{center}

\begin{abstract}
In the framework of the Lindblad theory for open quantum systems, we derive
closed analytical expressions of the Heisenberg and Schr\"odinger generalized
uncertainty functions for a particle moving in a harmonic oscillator potential.
The particle is initially in an arbitrary correlated coherent state, and interacts
with an environment at finite temperature. We describe how the quantum and
thermal fluctuations contribute to the uncertainties in the canonical variables of
the system and analyze the relative importance of these fluctuations in the
evolution of the system. We show that upon contact with the bath the system
evolves from a quantum-dominated to a thermal-dominated state in a time that
is of the same order as the decoherence time calculated in other models in the
context of  transitions from quantum to classical physics.
\end{abstract}


(a) e-mail address: isar@theory.nipne.ro

\section{Introduction}

The present work is directed towards the study of the Heisenberg and
Schr\"odinger  generalized uncertainty functions and the role of quantum and
thermal fluctuations during the evolution of a system consisting of a particle
moving in a harmonic oscillator potential and interacting with an environment.
There is a large amount of papers concerned with the question of modification
and generalization of the uncertainty principle, in particular on modified
uncertainty relations which incorporate the effect of environmentally induced
fluctuations \cite{AndH,AnH,afor}. The uncertainty relations derived in
\cite{AndH} are information-theoretic relations in terms of the Shannon
information (Wehrl entropy). In Ref. \cite{AnH}, Anastopoulos and Halliwell
derived such relations for the Heisenberg uncertainty function $
U=\sigma_{qq}\sigma_{pp}$ and for the Schr\"odinger generalized
uncertainty function $\sigma=\sigma_{qq}\sigma_{pp}-\sigma^2_{pq}.$
Here $\s_{qq},$ $\s_{pp}$ and  $\s_{pq}$ denote the variances and,
respectively, the covariance of coordinate and momentum of the open system.
In a series of papers Hu and Zhang computed the time evolution of the
Heisenberg uncertainty function $U$ in the presence of a thermal environment.
They noted for the first time the significance of the decoherence time scale for
quantum and thermal fluctuations of similar sizes \cite{hu1,hu2}. All these
authors used quantum Brownian models consisting of a particle moving in a
potential and linearly coupled to a bath of harmonic oscillators in a thermal
state. Their results show that in the Fokker-Planck regime, decoherence and
thermal fluctuations become important on the same time scale.

In this paper we consider Heisenberg and Schr\"odinger generalized
uncertainty functions for a harmonic oscillator coupled with an
environment. Our model is elaborated in the framework of the
Lindblad theory for open systems.

It is generally thought that quantum dynamical semigroups are the
basic tools to introduce dissipation in quantum mechanics
\cite{d,s,rev}. In Markovian approximation and for weakly damped
systems, the most general form of the generators of such
semigroups was given by Lindblad \cite{l1}. This formalism has
been studied extensively for the case of damped harmonic
oscillators \cite{rev,l2,ss,ass,a} and applied to various physical
phenomena, for instance, to the damping of collective modes in
deep inelastic collisions in nuclear physics \cite{i1}. A phase
space representation for the open quantum systems within the
Lindblad theory was given in Refs. \cite{i2,vlas}.

The paper is organized as follows. In Sec. 2 we remind the basic results
concerning the evolution of the damped harmonic oscillator in the Lindblad
theory for open quantum systems. Then in Sec. 3 we derive closed analytical
expressions for the finite temperature Heisenberg and Schr\"odinger
generalized uncertainty functions. We consider the general case of a thermal
bath and take the correlated coherent and squeezed states, in particular the
coherent states, as initial states. We consider the limiting cases of both zero
and high temperatures of the environment and the limit of short and long times.
In Sec. 4 we discuss the relative importance of quantum and thermal
fluctuations in the evolution of the system towards equilibrium with the aim of
clarifying the meaning of quantum, classical and thermal regimes, motivated by
the necessity of understanding the process of decoherence via interaction with
the environment and the general problem of the transition from quantum to
classical behaviour. In Sec. 5 we discuss our results in connection with other
work. A summary and concluding remarks are given in Sec. 6.

\section{Lindblad master equation for damped
harmonic oscillator}

The simplest dynamics for an open system which describes an
irreversible process is a semigroup of transformations introducing
a preferred direction in time \cite{d,s,l1}. In Lindblad's
axiomatic formalism of introducing dissipation in quantum
mechanics,
the irreversible time evolution of the open system
is described by the following general quantum Markovian master
equation for the density operator $\rho(t)$ \cite{l1}: \bea{d
\rho(t)\over dt}=-{i\over\hbar}[  H, \rho(t)]+{1\over 2\hbar}
\sum_{j}([  V_{j} \rho(t),  V_{j}^\dagger ]+[ V_{j}, \rho(t)
V_{j}^\dagger ]).\label{lineq}\eea Here $  H$ is the Hamiltonian
operator of the system and $  V_{j},$ $ V_{j}^\dagger $ are
operators on the Hilbert space
of the Hamiltonian, which model the environment. The semigroup
dynamics of the density operator which must hold for a quantum
Markovian process is valid only for the weak-coupling regime.
In the case of an exactly solvable model for the damped harmonic
oscillator, the two possible operators $ V_{1}$ and $ V_{2}$ are
taken as linear polynomials in coordinate $q$ and momentum $  p$
\cite{rev,l2,ss}
and the harmonic oscillator Hamiltonian $  H$ is chosen of the
most general quadratic form \bea
  H=  H_{0}+{\mu \over 2}(  q  p+  p  q),
~~~  H_{0}={1\over 2m}   p^2+{m\omega^2\over 2}  q^2. \label{ham}
\eea With these choices
the
master equation (\ref{lineq}) takes the following form
\cite{rev,ss}: \bea   {d \rho \over dt}=-{i\over \hbar}[ H_{0},
\rho]- {i\over 2\hbar}(\lambda +\mu) [  q, \rho   p+ p
\rho]+{i\over 2\hbar}(\lambda -\mu)[  p,
 \rho   q+  q \rho]  \nonumber\\
  -{D_{pp}\over {\hbar}^2}[  q,[  q, \rho]]-{D_{qq}\over {\hbar}^2}
[  p,[  p, \rho]]+{D_{pq}\over {\hbar}^2}([  q,[  p, \rho]]+ [ p,[
q, \rho]]). ~~~~\label{mast}   \eea The quantum diffusion
coefficients $D_{pp},D_{qq},$ $D_{pq}$ and the friction constant
$\lambda$ satisfy the following fundamental constraints
\cite{rev,ss}: $  D_{pp}>0, D_{qq}>0$ and \bea
D_{pp}D_{qq}-D_{pq}^2\ge {{\lambda}^2{\hbar}^2\over 4}.
\label{ineq}  \eea  In the particular case when the asymptotic
state is a Gibbs state $   \rho_G(\infty)=e^{-{  H\over kT}}/ {\rm
Tr}e^{-{  H\over kT}}, $ these coefficients can be written as
\cite{rev,ss} \bea D_{pp}={\lambda+\mu\over 2}\hbar
m\omega\coth{\hbar\omega\over 2kT}, ~~D_{qq}={\lambda-\mu\over
2}{\hbar\over m\omega}\coth{\hbar\omega\over 2kT}, ~~D_{pq}=0,
\label{coegib} \eea where $T$ is the temperature of the thermal
bath. In this case, the fundamental constraints are satisfied only
if $\la>\mu$ and \bea (\la^2-\mu^2)\coth^2{\hbar\om\over 2kT}
\ge\la^2.\label{cons}\eea We notice from this relation that in the
particular case of $T=0$ we must consider $\mu=0.$

Lindblad has proven \cite{l2} that in the Markovian regime the
harmonic oscillator master equation which satisfies the complete
positivity condition cannot satisfy simultaneously the
translational invariance and the detailed balance (which assures
an asymptotic approach to the canonical thermal equilibrium
state). The necessary and sufficient condition for translational
invariance is $\lambda=\mu$ \cite{rev,l2,ss}. In the following
general values $\lambda\neq \mu$ will be considered. In this way
we violate translational invariance, but we keep the canonical
equilibrium state.

By using the complete positivity property of the Lindblad model, in Refs.
\cite{rev,ss} the following inequality was obtained for all values of $t\ge 0$:
\bea
D_{pp}\sigma_{qq}(t)+D_{qq}\sigma_{pp}(t)-2D_{pq}\sigma_{pq}(t)\ge
{\hbar^2\lambda\over 2}.\label{has3}\eea The inequality (\ref{has3})
represents a restriction connecting the values of the variances and covariance
with the environment coefficients. In the case of a thermal bath, when the
environment coefficients have the form given by Eqs. (\ref{coegib}), the
condition (\ref{has3}) becomes \bea [(\lambda+\mu)
m\omega\sigma_{qq}(t)+(\lambda-\mu){\sigma_{pp}(t)\over
m\omega}]\coth{\hbar\omega\over 2kT}\ge \hbar\lambda \label{has5}.\eea
We have found in Ref. \cite{ass} that the inequality (\ref{has3}) is equivalent
with the generalized uncertainty relation at any time $t$
\bea\s(t)\equiv\sigma_{qq}(t)\sigma_{pp}(t)-\sigma_{pq}^2(t)\ge{\hbar^2
\over 4},\label{genun1}\eea if the initial values $\sigma_{qq}(0),\sigma_
{pp}(0)$ and $\sigma_{pq}(0)$ satisfy this inequality for $t=0$. The relation
(\ref{ineq}) is a necessary condition for the generalized uncertainty inequality
(\ref{genun1}) to be fulfilled. Schr\"odinger \cite{schr} and Robertson
\cite{rob} proved for the operators of coordinate $  q$ and momentum $  p$
that the inequality (\ref{genun1}) is in general fulfilled. The equality in relation
(\ref{genun1}) is realized for a special class of pure states, called correlated
coherent states \cite{dodkur} or squeezed coherent states.

The terms in Eq. (\ref{mast}) containing $\lambda$ and $\mu$ are
dissipative terms.
They cause a contraction of each volume element in phase space.
The diffusive terms containing $D_{pp}, D_{qq}$ and $D_{pq}$
produce an expansion of the volume elements and they are
responsible for noise and also for the destruction of interference
(decoherence).
Although the diffusion in general increases the uncertainty
$\s(t)$ as time goes on, there are competing effects that may
reduce it. In particular, wave packet reassembly (the time reverse
of wave packet spreading) and dissipation may cause the
uncertainty to decrease, but relation (\ref{genun1}) is always
obeyed.

From the master equation (\ref{mast}) we can obtain the equations of motion
for the variances and covariance of coordinate and momentum \cite{rev,ss}
which are needed to calculate the uncertainty functions:
\bea{d\sigma_{qq}(t)\over dt}=-2(\lambda-\mu)\sigma_{qq}(t)+{2\over m}
\sigma_{pq}(t)+2D_{qq},\eea \bea{d\sigma_{pp}\over
dt}=-2(\lambda+\mu)\sigma_{pp}(t)-2m\omega^2
\sigma_{pq}(t)+2D_{pp}, \label{eqmo2}\eea \bea{d\sigma_{pq}(t)\over
dt}=-m\omega^2\sigma_{qq}(t)+{1\over m}\sigma_{pp}(t)
-2\lambda\sigma_{pq}(t)+2D_{pq}.\eea In this paper we consider the
underdamped case $(\omega>\mu).$ Introducing the notations \bea
X(t)=\left(\matrix{m\omega\sigma_{qq}(t)\cr \sigma_{pp}(t)/m\omega\cr
\sigma_{pq}(t)\cr}\right),~~~~\label{var} D=\left(\matrix{2m\omega
D_{qq}\cr 2D_{pp}/m\omega\cr 2D_{pq}\cr}\right),\eea the solutions of
these equations of motion can be written in the form \cite{rev,ss} \bea
X(t)=(Te^{Kt}T)(X(0)-X(\infty))+X(\infty),\label{sol2}\eea where the
matrices $T$ and $K$ are given by (we introduce the notation
$\Omega^2=\omega^2-\mu^2$) \bea T={1\over
2i\Omega}\left(\matrix{\mu+i\Omega&\mu-i\Omega&2\omega\cr
\mu-i\Omega&\mu+i\Omega&2\omega\cr
-\omega&-\omega&-2\mu\cr}\right),~~~~
K=\left(\matrix{-2(\lambda-i\Omega)&0&0\cr 0&-2(\lambda+i\Omega)&0\cr
0&0&-2\lambda\cr}\right)\label{matr}\eea and \bea
X(\infty)=-(TK^{-1}T)D.\label{xinf}\eea Formula (\ref{xinf}) gives a simple
connection between the asymptotic values $(t \to \infty)$ of $\sigma_{qq}(t),
\sigma_{pp}(t), \sigma_{pq}(t)$ and the diffusion coefficients
$D_{pp},D_{qq},D_{pq}.$ These asymptotic values do not depend on the
initial values $\sigma_{qq}(0),$ $ \sigma_{pp}(0),$ $\sigma_{pq}(0)$ and in
the case of a thermal bath with coefficients (\ref{coegib}), they reduce to
\cite{rev,ss} \bea \s_{qq}(\infty)={\hbar\over
2m\om}\coth{\hbar\omega\over 2kT}, ~~\s_{pp}(\infty)={\hbar
m\omega\over 2}\coth{\hbar\omega\over 2kT}, ~~\s_{pq}(\infty)=0.
\label{varinf} \eea

\section{Heisenberg and Schr\"odinger uncertainty functions for
Lindblad model}

\subsection{Choice of initial wave function}

We consider a harmonic oscillator with an initial wave function
\bea \Psi(x)=({1\over 2\pi\sigma_{qq}(0)})^{1\over 4}\exp[-{1\over
4\sigma_{qq}(0)}
(1-{2i\over\hbar}\sigma_{pq}(0))(x-\sigma_q(0))^2+{i\over
\hbar}\sigma_p(0)x], \label{ccs}\eea where $\s_{qq}(0)$ is the
initial spread, $\s_{pq}(0)$ the initial covariance, and $\s_q(0)$
and $\s_p(0)$ are the averaged initial position and momentum of
the Gaussian wave packet. As will be seen, the parameters $\s_q$
and $\s_ p$ do not appear in the uncertainty functions.
As initial state we take a correlated coherent state \cite{dodkur}
which is represented by the Gaussian wave packet (\ref{ccs}) in
the coordinate representation with the variances and covariance of
coordinate and momentum \bea \s_{qq}(0)={\hbar\de\over 2m\om},~~
\s_{pp}(0)={\hbar m\om\over 2\de(1-r^2)},~~ \s_{pq}(0)={\hbar
r\over 2\sqrt{1-r^2}}. \label{inw}\eea Here, $\de$ is the
squeezing parameter which measures the spread in the initial
Gaussian packet and $r=r(0),$ $|r|<1$ is the correlation
coefficient at time $t=0.$ The correlation coefficient is defined
as \bea
r(t)={\sigma_{pq}(t)\over\sqrt{\sigma_{qq}(t)\sigma_{pp}(t)}}.
\label{corcoe}\eea The initial values (\ref{inw}) correspond to a
so-called
minimum uncertainty state, since they fulfil the generalized
uncertainty relation with equal sign \bea \s(0)\equiv
\sigma_{qq}(0)\sigma_{pp}(0)-\sigma_{pq}^2(0) ={\hbar^2\over
4}.\label{gen0}\eea For $\de=1$ and $r=0$ the correlated coherent
state becomes a Glauber coherent state. With the initial values
(\ref{inw}) and by assuming a thermal bath with temperature $T,$
the condition (\ref{has5}) for $t=0$ takes the form
($\epsilon\equiv\hbar\omega/2kT$):
\bea [(\lambda+\mu)\de+(\lambda-\mu){1\over
\de(1-r^2)}]\coth\epsilon\ge 2\lambda \label{has7}.\eea One can
easily show that \bea [(\lambda+\mu)\de+(\lambda-\mu){1\over
\de}]\ge 2\sqrt{\lambda^2-\mu^2}.\eea It follows that if relation
(\ref{cons}) is satisfied, then relation (\ref{has7}) is also
satisfied (since $|r|<1$) and, therefore, for a given temperature
$T$ of the bath and for any parameters $\delta$ and $r$ the
inequality (\ref{cons}) alone determines the range of values of
the parameters $\lambda$ and $\mu.$

\subsection{Heisenberg uncertainty function at finite temperature}

For simplicity we set $r=0$ in this Subsection. With the variances
given by Eq. (\ref{sol2})
we calculate the  Heisenberg uncertainty function
$U(t)=\sigma_{qq}(t)\sigma_{pp}(t)$ for finite
temperature and obtain: \bea
U(t)={\hbar^2\over 4}\{e^{-4\la
t}[1-(\de+{1\over\de})\coth\epsilon+\coth^2\epsilon+{\om^2\over
4\Om^4}[\om^2(\de-{1\over\de})^2\sin^2(2\Om t)\nonumber \\
+2\mu^2
[(\de-\coth\epsilon)^2+({1\over\de}-\coth\epsilon)^2]\cos(2\Om t)(\cos(2\Om t)-1)\nonumber\\
+4\mu^2(\de-\coth\epsilon)({1\over\de}-\coth\epsilon)(1-\cos(2\Om t))\nonumber\\
+2\mu\Om[(\de-\coth\epsilon)^2-({1\over\de}-\coth\epsilon)^2]\sin(2\Om
t)(1-\cos(2\Om
t))]]\nonumber\\
+e^{-2\la
t}\coth\epsilon[(\de+{1\over\de}-2\coth\epsilon){\om^2-\mu^2\cos(2\Om
t)\over \Om^2} +(\de-{1\over\de}){\mu\sin(2\Om
t)\over\Om}]+\coth^2\epsilon \}.\label{genunc}\eea This is the first main
result of the present paper.

For an initial
coherent state ($\de=1$), this expression simplifies: \bea
U(t)={\hbar^2\over 4}\{e^{-4\la
t}(\coth\epsilon-1)^2[1+{\om^2\mu^2\over \Om^4}(1-\cos(2\Om t))^2]\nonumber\\
+2e^{-2\la t}\coth\epsilon(1-\coth\epsilon){\om^2-\mu^2\cos(2\Om
t)\over \Om^2}+\coth^2\epsilon \}.\eea
For $\mu=0,$ which is a more natural choice, since it corresponds
to the usual form of the oscillator Hamiltonian (\ref{ham}), we
obtain from Eq. (\ref{genunc}): \bea U(t)={\hbar^2\over
4}\{e^{-4\la t}[1-(\de+{1\over\de})\coth\epsilon+\coth^2\epsilon
+{1\over 4}(\de-{1\over\de})^2\sin^2(2\om t)]\nonumber\\
+e^{-2\la t}\coth\epsilon(\de+{1\over\de}-2\coth\epsilon)+\coth^2\epsilon
\}.\label{hunc}\eea For $\de=1$ this expression takes the form \bea
U(t)={\hbar^2\over 4}\{e^{-2\la t}+\coth\epsilon(1-e^{-2\la
t})\}^2.\label{gen1} \eea Here the first term is of quantum nature, whereas
the second term is of thermal nature. Their contributions to the uncertainty of
the system arise from quantum and thermal fluctuations, respectively. This
expression is similar to that obtained in Refs. \cite{hu1,hu2}, the difference
being that in $\coth\epsilon$ instead of the natural frequency $\omega$ stays
the renormalized frequency.

In the obtained expressions of the uncertainty, besides the
parameters $\lambda,\mu$ and $\de$ there are also two factors,
time and temperature, which we will consider for different
regimes. For example, in the case of $T=0$ $(\coth\epsilon=1)$ we have to
take, according to Eq. (\ref{cons}), $\mu=0$ and then
we obtain from Eq. (\ref{hunc}):
\bea U_0(t)= {\hbar^2\over 4}\{1+e^{-4\la t}[2-(\de+{1\over\de})
+{1\over 4}(\de-{1\over\de})^2\sin^2(2\om t)] +e^{-2\la
t}[(\de+{1\over\de}-2)\}.\label{unzer}\eea We see in the last
expression that the leading term is given by $\hbar^2/4$ (the
Heisenberg contribution) followed, for squeezed states $\de\neq
1,$ by terms describing both decay and oscillatory behaviour,
representing quantum fluctuations alone (since $T=0).$ For $\de=1$
we obtain in the zero-temperature case $U_0(t)=\hbar^2/4.$

\subsection{Schr\"odinger uncertainty function at finite temperature}

With the variances given by Eq. (\ref{sol2})
we calculate now the finite temperature generalized uncertainty
function $\s(t)=\sigma_{qq}(t)\sigma_{pp}(t)-\sigma_{pq}^2(t)$ and
obtain: \bea \s(t)={\hbar^2\over 4}\{e^{-4\la
t}[1-(\de+{1\over\de(1-r^2)})\coth\epsilon+\coth^2\epsilon]
\nonumber \\
+e^{-2\la
t}\coth\epsilon[(\de+{1\over\de(1-r^2)}-2\coth\epsilon){\om^2-\mu^2\cos(2\Om
t)\over\Om^2}\nonumber \\
+(\de-{1\over\de(1-r^2)}){\mu \sin(2\Om t)\over\Om}+{2r\mu\om
(1-\cos(2\Om
t))\over\Om^2\sqrt{1-r^2}}]+\coth^2\epsilon\}.\label{sunc}\eea This is the
second main result of the present paper.

When the correlation coefficient $r=0,$ the correlated coherent
initial state becomes a pure squeezed state. In this case the
expression of the uncertainty function takes the form \bea
\s(t)={\hbar^2\over 4}\{e^{-4\la
t}[1-(\de+{1\over\de})\coth\epsilon+\coth^2\epsilon]
\nonumber \\
+e^{-2\la
t}\coth\epsilon[(\de+{1\over\de}-2\coth\epsilon){\om^2-\mu
^2\cos(2\Om t)\over \Om ^2}+(\de-{1\over\de}){\mu \sin(2\Om
t)\over\Om}]+\coth^2\epsilon\}.\label{sunc1}\eea

For the usual coherent state ($\de=1$) the uncertainty function
is  \bea \s(t)={\hbar^2\over 4}\{e^{-4\la
t}(\coth\epsilon-1)^2
\nonumber \\
+2e^{-2\la t}\coth\epsilon(1-\coth\epsilon){\om^2-\mu ^2\cos(2\Om
t)\over \Om^2}+\coth^2\epsilon\}.\eea

All the expressions above simplify in the particular case when
$\m=0.$ Namely, in this case the uncertainty function (\ref{sunc})
has the form \bea \s(t)={\hbar^2\over 4}\{e^{-4\la
t}[1-(\de+{1\over\de(1-r^2)})\coth\epsilon+\coth^2\epsilon]
\nonumber \\
+e^{-2\la
t}\coth\epsilon[\de+{1\over\de(1-r^2)}-2\coth\epsilon]+\coth^2\epsilon\}.\eea
For an initial squeezed state $(r=0)$ and any squeezing
coefficient $\de$, we obtain \bea \s(t)={\hbar^2\over 4}\{e^{-4\la
t}[1-(\de+{1\over\de})\coth\epsilon+\coth^2\epsilon]
\nonumber \\
+e^{-2\la
t}\coth\epsilon(\de+{1\over\de}-2\coth\epsilon)+\coth^2\epsilon\}.
\label{sunc2}\eea
When the initial state is the usual coherent state $(\de=1)$, we
obtain \bea \s(t)={\hbar^2\over 4}\{e^{-2\la
t}+\coth\epsilon(1-e^{-2\la t})\}^2,\eea which is identical with
Eq. (\ref{gen1})
for $U(t).$

We consider now the particular case when the temperature of the
thermal bath is $T=0.$ Then we have to set also $\mu=0$ (cf. Eq.
(\ref{cons})) and the uncertainty function $\s(t)$ takes the
following form:
\bea \s_0(t)={\hbar^2\over 4}\{1+(e^{-4\la t}-e^{-2\la t})
[2-(\de+{1\over\de(1-r^2)})]\}.\label{sizer}\eea

We see that in the last expression the leading term is given by $\hbar^2/4$
(the Heisenberg contribution) followed by terms representing quantum
fluctuations alone (since $T=0$). Compared to Eq. (\ref{unzer}), where these
terms  describe both decay and oscillating behaviour, in Eq. (\ref{sizer}) the
terms representing the quantum fluctuations describe only a decay behaviour.
When the initial state is the usual coherent state $(\de=1, r=0)$, the
uncertainty function takes again the most simple form $\s_0(t)=\hbar^2/4$ for
all times.

\section{Transition from quantum mechanics to classical
statistical mechanics}

~~~~~~(a) $t=0:$ When the initially uncorrelated condition is assumed valid,
we have $\s(0)=U(0)=\hbar^2/4,$ according to Eq. (\ref{gen0}).

(b) $t\gg \la^{-1}$ (very long times): $\s(t)$ and $U(t)$ are insensitive to
$\la,\mu,\de$ and $r$ and approach \bea \s^{BE}=U^{BE}={\hbar^2\over
4}\coth^2\epsilon,\label{ube}\eea which is a Bose-Einstein relation for a
system of bosons in equilibrium at temperature $T$ (quantum statistical
mechanics). Again $T=0$ is the limit of pure quantum fluctuations, \bea
\s_0=U_0={\hbar^2\over 4},\label{hut}\eea which is the quantum
Heisenberg relation and high $T$ $(T\gg \hbar\om/k)$ is the limit of pure
thermal fluctuations, \bea
\s^{MB}=U^{MB}=({kT\over\om})^2,\label{umb}\eea which is a
Maxwell-Boltzmann distribution for a system approaching a classical limit (in
classical statistical mechanics the equipartition theorem imparts for each degree
of freedom an uncertainty of $kT/2$ ). These are expected results from
quantum mechanics and classical statistical mechanics. The formula (\ref{ube})
interpolates between the two results (\ref{hut}) at $T=0$ and (\ref{umb}) at
$T\gg\hbar\om/k.$

(c) $r=0:$ At short times $(\la t\ll 1, \Om t\ll 1),$ we obtain from Eqs.
(\ref{genunc}) and (\ref{sunc1}): \bea \s(t)=U(t)= {\hbar^2\over
4}\{1+2[\la
(\de+{1\over\de})\coth\epsilon+\mu(\de-{1\over\de})\coth\epsilon-
2\la]t\}.\label{sho1}\eea The time when thermal fluctuations overtake
quantum fluctuations is \bea t_d={1\over 2[\la
(\de+{1\over\de})\coth\epsilon+\mu(\de-{1\over\de})\coth\epsilon-2\la]}.
\label{t1}\eea According to the theory of Halliwell \cite{AndH,AnH} and Hu
\cite{hu1,hu2}, we expect this time to be equal to the decoherence time scale,
which is not yet calculated for the damped harmonic oscillator in the Lindblad
model for open quantum systems.

For $\mu=0$ we obtain from Eq. (\ref{sho1}): \bea \s(t)=U(t)=
{\hbar^2\over 4}\{1+2\la
[(\de+{1\over\de})\coth\epsilon-2]t\}.\eea  In this case \bea
t_d={1\over 2\la [(\de+{1\over\de})\coth\epsilon-2]}.\eea For
$\de=1$ the uncertainty function (\ref{sho1}) is independent of
$\mu.$

(i) At temperature $T=0$ the uncertainty (\ref{sho1}) becomes $(\mu=0)$
\bea \s_0(t)=U_0(t)= {\hbar^2\over 4}\{1+2\la (\de+{1\over\de}-2)t\}\eea
and \bea t_d={1\over 2\la (\de+{1\over\de}-2)}.\eea

(ii) In the case of high temperatures, introducing the notation \bea \tau\equiv
{2kT\over \hbar\omega}\equiv {1\over \epsilon},\eea we obtain \bea
\s(t)=U(t)= {\hbar^2\over 4}\{1+2[\la
(\de+{1\over\de})\tau+\mu(\de-{1\over\de})\tau-2\la]t\}\label{sho2} \eea
and the time when thermal fluctuations overtake quantum fluctuations is given
by \bea t_d={\hbar\omega\over 4kT[\la
(\de+{1\over\de})\tau+\mu(\de-{1\over\de})\tau]}.\label{sho3}\eea

For $\de=1$ we obtain from Eq. (\ref{sho2}) \bea \s(t)=U(t)={\hbar^2\over
4}\{1+4\la (\tau-1)t\},\eea independent of $\mu$ and \bea
t_d={\hbar\omega\over 8kT\la}.\eea This particular result coincides with those
obtained in Refs. \cite{hu1,hu2}.

(d) $r\neq 0:$ In the rest of this Section we analyze the generalized uncertainty
function $\s$ in the case when the correlation coefficient $r$ is different from 0.
At short times $(\la t\ll 1, \Om t\ll 1),$ we obtain from Eq. (\ref{sunc}): \bea
\s(t)= {\hbar^2\over 4}\{1+2[\la
(\de+{1\over\de(1-r^2)})\coth\epsilon+\mu(\de-{1\over\de(1-r^2)})
\coth\epsilon- 2\la]t\}.\label{sho4}\eea The time when thermal fluctuations
become comparable with quantum fluctuations is in this case \bea t_d={1\over
2[\la (\de+{1\over\de(1-r^2)})\coth\epsilon+\mu(\de-{1\over\de(1-r^2)})
\coth\epsilon-2\la]}.\label{t2}\eea

(i) At zero temperature $T=0,$ the uncertainty becomes ($\mu=0$): \bea
\s_0(t)= {\hbar^2\over 4}\{1+2\la (\de+{1\over\de(1-r^2)}-2)t\}\eea and
\bea t_d={1\over 2\la (\de+{1\over\de(1-r^2)}-2)}.\eea

(ii) At high temperature \bea \s(t)= {\hbar^2\over 4}\{1+2[\la
(\de+{1\over\de(1-r^2)})\tau+\mu(\de-{1\over\de(1-r^2)})\tau
-2\la]t\}\eea and the time when thermal fluctuations overtake quantum
fluctuations is given by \bea t_d={\hbar\omega\over 4kT[\la
(\de+{1\over\de(1-r^2)})\tau+\mu(\de-{1\over\de(1-r^2)})\tau]}.\eea

In summary of this section, the third main result is represented by: (1) the
expresions (\ref{sho1}) of the uncertainty $\sigma=U$ (for $r=0$) and
(\ref{sho4}) of the uncertainty $\sigma$ (for $r\neq 0$) for short initial times,
which evidently fulfil the uncertainty principle by virtue of the condition
(\ref{has7}); (2) the corresponding expressions (\ref{t1}) and (\ref{t2}) for
the time when the thermal fluctuations become comparable with the quantum
fluctuations; we notice that this time is decreasing with the increasing of both
temperature $T$ and dissipation $\lambda$.

\section{Discussion of results}

One often regards the regime where thermal fluctuations begin to surpass
quantum fluctuations as the transition point from quantum to classical statistical
mechanics and identifies the high temperature regime of a system as the
classical regime.  On the other hand, it is known that a necessary condition for
a system to behave classically is that the interference terms in its wave function
have to diminish below a certain level, so that probability can be assigned to
classical events \cite{hu1,hu2}. This is the decoherence process. The
decoherence via interaction with an environment views the disappearance of the
off-diagonal components of a reduced density matrix in some special basis as
signaling a transition from quantum to classical physics. In Refs.
\cite{AndH,AnH,hu1,hu2} it was shown that these two criteria of classicality are
equivalent: the time when the quantum system decoheres is also the time when
thermal fluctuations overtake quantum fluctuations. However the regime after
thermal fluctuations dominate should not be called classical. After the
decoherence time, although the system is describable in terms of probabilities,
it cannot yet be regarded as classical because of the spin-statistics effects and
has to be described by non-equilibrium quantum statistical mechanics. Only
after the relaxation time the system can be correctly described by the
equilibrium quantum statistical mechanics. The classical regime starts at a much
later time. Only at a sufficiently high temperature when the spin (Fermi-Dirac or
Bose-Einstein) statistics can be represented by the Maxwell-Boltzmann
distribution function, can the system be considered in a classical regime
\cite{hu1,hu2} (see Eqs. (\ref{ube}) -- (\ref{umb})).

The case of zero coupling, $\la=0$ and  $\mu=0,$ corresponds to an isolated
harmonic oscillator taken as a closed quantum system. We find the Heisenberg
quantum uncertainty function for an initial squeezed state to be (see Eq.
(\ref{hunc})) \bea U(t)= {\hbar^2\over 4}\{1+{1\over 4}
(\de-{1\over\de})^2\sin^2(2\om t)\}\ge {\hbar^2\over 4}.\eea This is the
quantum uncertainty relation for squeezed states. As the coupling to the
environment goes to 0, the thermal fluctuations go also to 0 and the
time-dependent term is the result of quantum fluctuations only. For the
unsqueezed coherent state, $\de=1,$ we recover the Heisenberg uncertainty
relation $U(t)={\hbar^2/ 4}.$ For the same case of zero coupling, the
Schr\"odinger uncertainty function (\ref{sunc}) becomes $ \s(t)={\hbar^2/
4}$ for any correlated coherent initial state.

In all expressions for the uncertainty functions obtained in the preceding Sec. 4,
the terms depending on $t$ are functions of the initial spread and correlation
coefficient and represent the initial growth of thermal fluctuations, starting from
the pure quantum fluctuations at $t=0.$ Using condition (\ref{has7}), we
notice that for short initial times the uncertainty increases with dissipation $\la$
and temperature $T.$ This is in contrast with other models studied in literature
\cite{AndH,AnH,hu1,hu2}, where the uncertainty principle is violated  on a
short time scale as a consequence of the well-known violation of the positivity
of the density operator \cite{AnH,cald,cald1,amb}. For instance, in the model
considered in Ref. \cite{AnH}, in the Fokker-Planck (high temperature) limit the
uncertainty principle is violated for times $t<(h^2/\gamma k^2T^2)^{1/3}$
in the case of uncertainty $\s$ and for times $t<\hbar/kT$ in the case of
uncertainty $U$. Likewise, the uncertainty function $U$ obtained in Refs.
\cite{hu1,hu2} for high temperature $T$ satisfies the uncertainty principle only
for $t>\hbar/kT.$ In addition, for a finite temperature there exists a restriction
on how low the temperature could become for a given squeezing parameter, in
order that the uncertainty principle to be satisfied \cite{hu2}. In fact, the
uncertainty principle in our model is fulfilled not only for short times, but for any
time and temperature and for the full range of the squeezing and correlation
parameters. Indeed, cf. Eq. (\ref{genun1}), by virtue of the complete positivity
property of the Lindblad model, the generalized uncertainty function $\s$
always fulfills the uncertainty principle and for $U(\ge\s)$ this is also true.

To exemplify the evolution of the uncertainty functions, in Fig. 1 we represent
the Heisenberg $U$ and Schr\"odinger $\s$ uncertainties as functions on time
and on temperature via $\coth\epsilon,$ for $\mu=0, r=0$ and for a given
value of the squeezing parameter. In Fig. 2 the same uncertainties are
represented as functions on time and on squeezing parameter, for $\mu=0,
r=0$ and for a finite temperature of the environment. In Fig. 3 the uncertainties
are represented also as functions on time and on temperature for $r=0$ and for
given values of the squeezing parameter $\de$ and $\mu\neq 0.$  In Fig. 4 we
represent the dependence of the uncertainty function $\s$ on time, on
temperature and on the squeezing parameter for given finite values of $r$ and
$\mu.$

\begin{figure}
\label{Fig. 1} \centerline{\epsfig{file=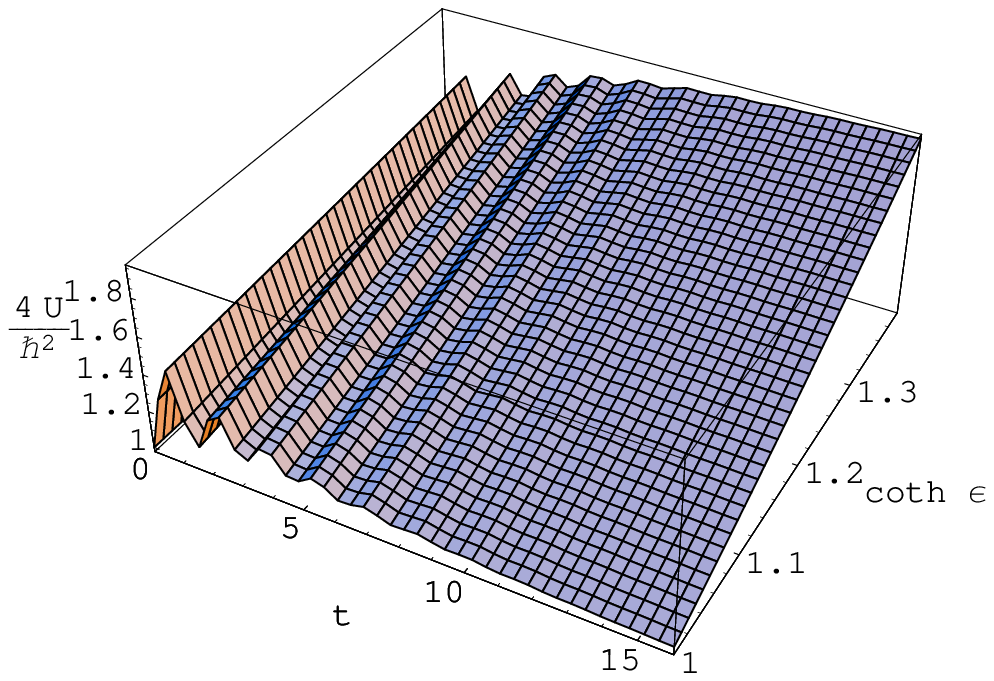, width=0.6\textwidth}}
\centerline{a)}
 \centerline{\epsfig{file=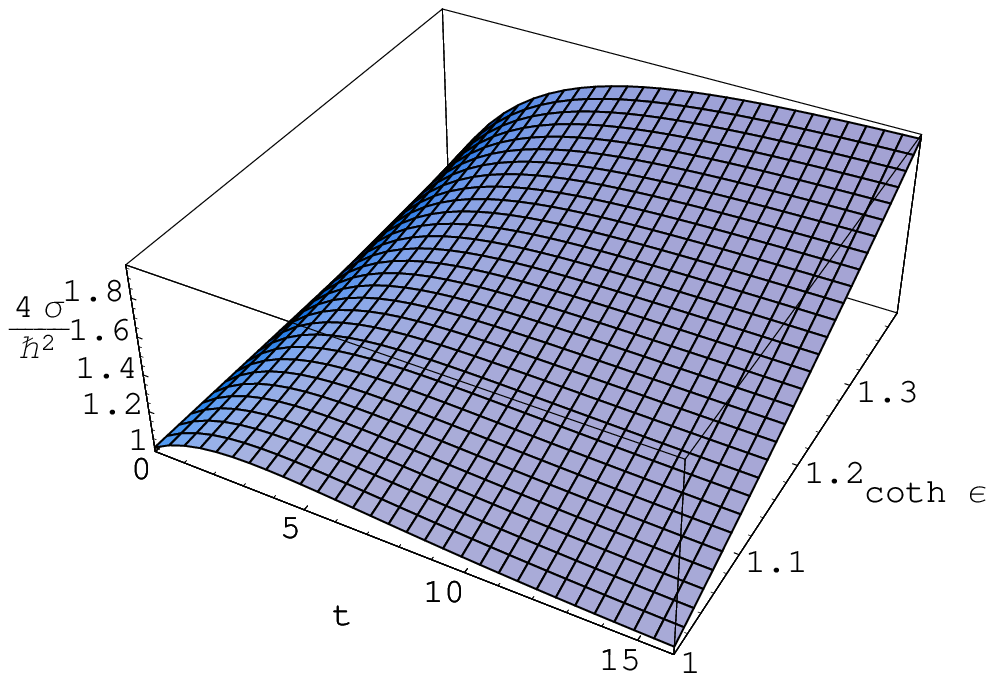, width=0.6\textwidth}}
\centerline{b)} \caption{a) Dependence of the Heisenberg uncertainty $U$
(\ref{hunc}) on time $t$ and on temperature $T$ via $\coth\epsilon$
($\epsilon\equiv\coth({\hbar\omega/2kT})),$ for $\omega=1,$
$\lambda=0.1,$ $\mu=0,$ $r=0$ and $\de=2.$ b) Dependence of the Schr\"
odinger uncertainty $\s$ (\ref{sunc2}) on the same variables and for the same
values of the parameters. The unit of time is $s.$ }
\end{figure}

\begin{figure}
\label{Fig. 2} \centerline{\epsfig{file=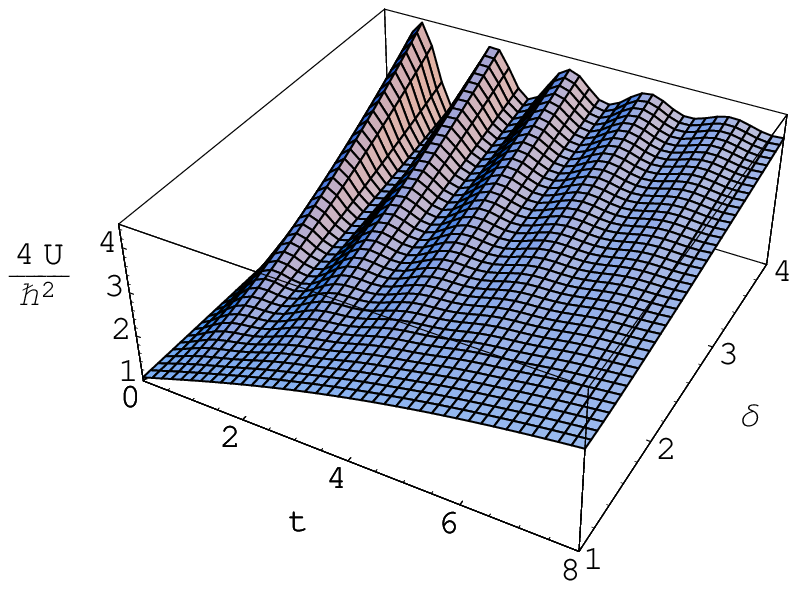, width=0.6\textwidth}}
\centerline{a)} \centerline{\epsfig{file=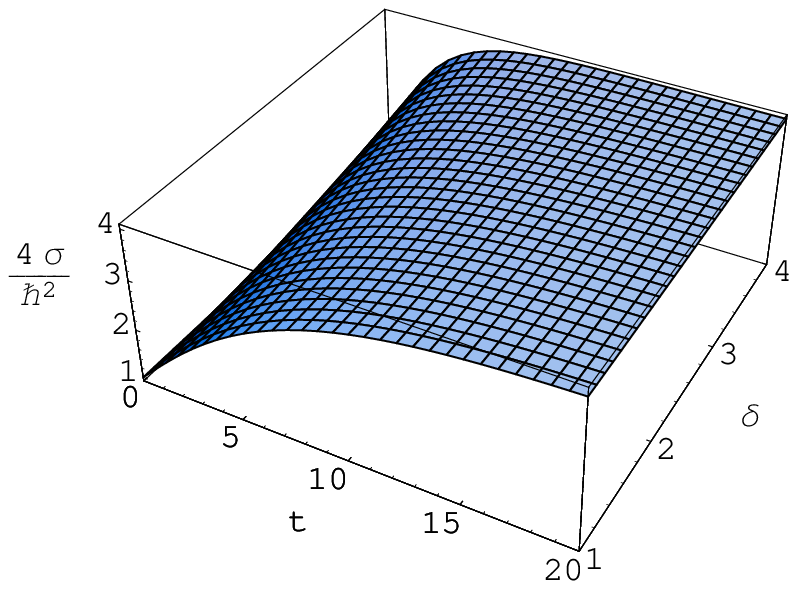, width=0.6\textwidth}}
\centerline{b)} \caption{a) Dependence of the Heisenberg uncertainty $U$
(\ref{hunc}) on time $t$ and on the squeezing parameter $\de$ for
$\omega=1,$ $\lambda=0.1,$ $\mu=0,$ $r=0$ and $\coth\epsilon=2.$ b)
Dependence of the Schr\"odinger uncertainty $\s$ (\ref{sunc2}) on the same
variables and for the same values of the parameters. The unit of time is $s.$}
\end{figure}

\begin{figure}
\label{Fig. 3} \centerline{\epsfig{file=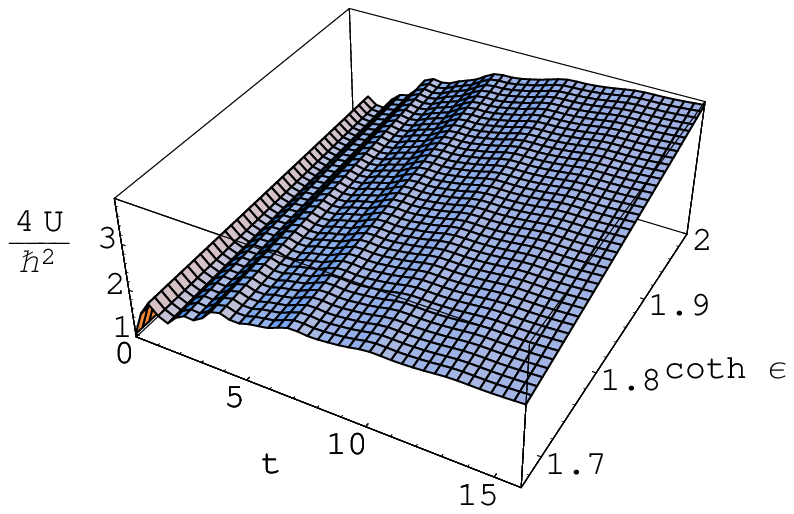, width=0.6\textwidth}}
\centerline{a)} \centerline{\epsfig{file=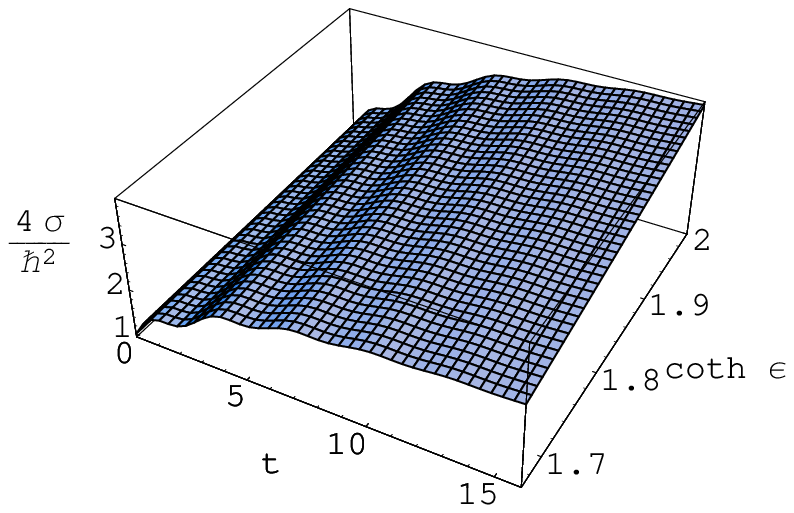, width=0.6\textwidth}}
\centerline{b)} \caption{a) Dependence of the Heisenberg uncertainty $U$
(\ref{genunc}) on time $t$ and on temperature $T$ via $\coth\epsilon$
($\epsilon\equiv\coth({\hbar\omega/2kT})),$ for $\omega=1,$
$\lambda=0.1,$ $\mu=0.08,$ $r=0$ and $\de=2.$ b) Dependence of the
Schr\" odinger uncertainty $\s$ (\ref{sunc1}) on the same variables and for
the same values of the parameters. The unit of time is $s.$}
\end{figure}

\begin{figure}
\label{Fig. 4} \centerline{\epsfig{file=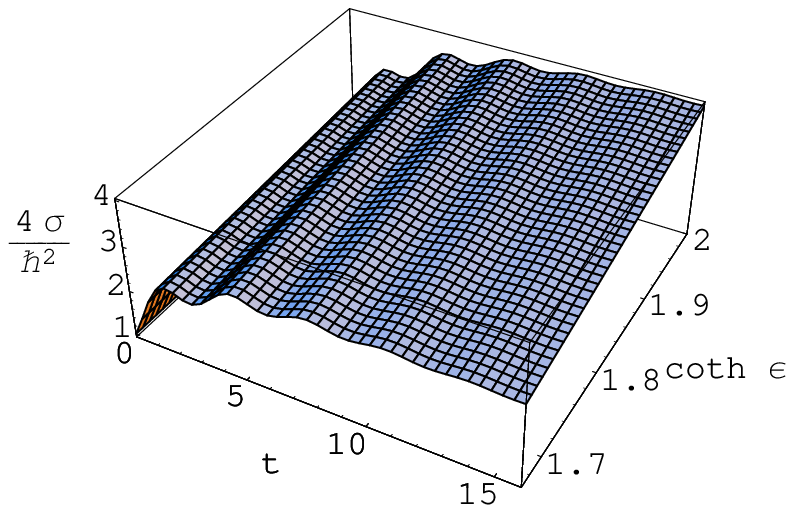, width=0.6\textwidth}}
\centerline{a)} \centerline{\epsfig{file=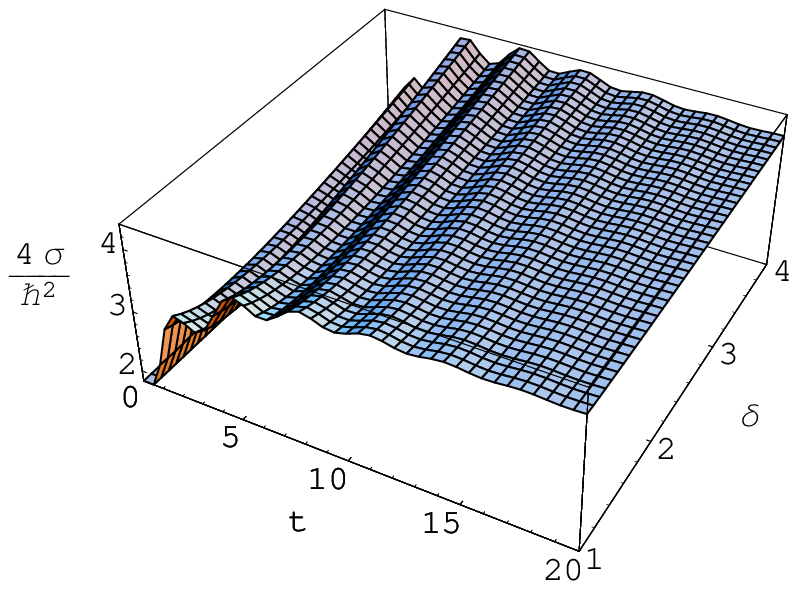, width=0.6\textwidth}}
\centerline{b)} \caption{a) a) Dependence of the Schr\"odinger uncertainty
$\s$ (\ref{sunc}) on time $t$ and on temperature $T$ via $\coth\epsilon$
($\epsilon\equiv\coth({\hbar\omega/2kT})),$ for $\omega=1,$
$\lambda=0.1,$ $\mu=0.08,$ $r=0.8$ and $\de=2.$ b) Dependence of the
same Schr\"odinger uncertainty $\s$ on time $t$ and on the squeezing
parameter $\de$ for $\omega=1,$ $\lambda=0.1,$ $\mu=0.08,$ $r=0.8$ and
$\coth\epsilon=2.$ The unit of time is $s.$}
\end{figure}

The time dependence of the uncertainties $U(t)$ and
$\s(t)$ given by the dissipative terms reflects the fact
that the Lindblad evolution of the system is non-unitary
and is an expression of the effect of the environment.
This is in contrast with the usual Liouville-von Neumann
unitary evolution, when the uncertainty is independent of
time, being invariant under unitary transformations.

The limiting case $\la=\mu$ deserves special attention in the Lindblad theory.
As we mentioned in Sec. 2, the translational invariance is fulfilled in this case,
but the asymptotic state is not a thermal state. Formally, the master equations
considered by Halliwell and Hu \cite{AndH,AnH,hu1,hu2} can be obtained by
taking $D_{qq}=0$ in the Lindblad master equation. In Lindblad theory this
choice is forbidden, since in this case the fundamental constraints (\ref{ineq})
are not any more fulfilled and this is the fact connected with the violation of the
positivity of the density matrix in the mentioned models.

The time $t_d$ when thermal fluctuations overtake quantum fluctuations
obtained in our model for different initial states of the open system (correlated
coherent states, squeezed states, coherent states) has in general a different
form compared to the corresponding time obtained in Refs.
\cite{AndH,AnH,hu1,hu2}. In the high temperature limit it is of the same scale
as the decoherence time (time when the off-diagonal components of the density
matrix responsible for interference decrease to zero due to the interaction with
the environment). The value of this time was determined in a series of papers in
quantum Brownian motion models for initial coherent states
\cite{blhu,paz1,paz2,zur}, but not yet  in the Lindblad model.

The second time scale of importance is the relaxation time scale,
$t_{rel}=\la^{-1}\gg t_d,$ when the particle reaches equilibrium with the
environment.  After this time, the uncertainty function takes on the
Bose-Einstein form (\ref{ube}). At high temperatures the system reaches the
Maxwell-Boltzmann limit and the uncertainty function takes on the classical form
(\ref{umb}).

In the case of zero temperature, there are no longer thermal fluctuations and
the environmentally induced fluctuations are of quantum nature only, given by
terms describing both decay and oscillatory behaviour in the case of the
uncertainty function $U$ (\ref{unzer}) and only decay behaviour in the case of
the uncertainty function $\sigma$ (\ref{sizer}).

Many of the present results are related to those obtained in the cited papers
\cite{AndH,AnH,hu1,hu2} and they lead essentially to the same physical
conclusions in what concerns the role of thermal and quantum fluctuations in
the transition from quantum to classical behaviour, in concordance with the
results of quantum mechanics, quantum and classical statistical mechanics. In
this connection, we would like to add the following remarks concerning the
differences between our paper and those earlier works.

(1) The uncertainty relations are studied in Refs. \cite{AndH,AnH,hu1,hu2} in
the quantum Brownian motion model of an open system consisting of a
harmonic oscillator linearly coupled to a thermal bath, using the
Feynman-Vernon influence functional formalism to incorporate the statistical
effect of the environment on the system in the reduced density matrix. In our
paper, we use the theory of open quantum system based on the axiomatic
approach of completely positive dynamical semigroups, which gives the most
general form of a quantum Markovian master equation for the density operator
of the open system.

(2) As is well-known, in the theory based on quantum dynamical semigroups
the density operator satisfies the necessary requirements of unitarity,
Hermiticity and positivity, while in the quantum Brownian motion model the
positivity of the reduced density matrix is violated at short time scale. As a
consequence, the uncertainty relations obtained in Refs.
\cite{AndH,AnH,hu1,hu2} for both Heisenberg $U$ and Schr\"odinger $\s$
uncertainties violate the uncertainty principle for short initial time, for all
temperatures $T$, including the Fokker-Planck limit of high temperatures. The
only relation which fulfills the uncertainty principle for all $t$ is obtained in Ref.
\cite{AnH}, with the expense that it contains the finite cutoff frequency
introduced in the spectral density. As we already mentioned, in the Lindblad
model both uncertainty functions $U$ and $\s$ satisfy the uncertainty principle
for the whole range of time $t,$ temperature $T,$ dissipative constant
$\lambda$ and for any initial correlated coherent state determined by the
squeezing parameter $\delta$ and correlation coefficient $r.$

(3) In general, the expressions of the time on which the thermal fluctuations
become comparable to the quantum ones are obtained in Refs.
\cite{AndH,AnH,hu1,hu2} in the Fokker-Planck limit of high temperatures. For
the case of general $T,$ even in the only expression obtained in Refs.
\cite{hu1,hu2} for this time, a restriction has to be imposed on how low $T$
could become for a given squeezing parameter. The expressions from the
present paper for the time on which thermal fluctuations overtake the quantum
ones are obatined for any temperature and contain an explicit dependence on
$\lambda,\delta$ and $r.$ They have a different form from those obtained in
the cited works (only in the particular case of high temperatures and for an
initial coherent state $(\delta=1,r=0)$, our result is identical to that obtained in
Refs. \cite{hu1,hu2}) and in the high temperature limit they are comparable
with the decoherence time calculated in decoherence studies of transitions from
quantum to classical behaviour.

\section{Summary and concluding remarks}

In the present paper we have studied the evolution of the one-dimensional
harmonic oscillator with dissipation within the framework of the Lindblad theory
for open quantum systems. We have considered the general case of an
environment consisting of a thermal bath at an arbitrary temperature. The
Gaussian correlated coherent, squeezed and Glauber coherent states were
taken as initial states. We have derived closed analytical expressions of the
Heisenberg and Schr\"odinger uncertainty functions for the evolution of the
damped harmonic oscillator for different regimes of time and temperature, in
particular in the limiting cases of both zero temperature and high temperature
of the environment, in the limit of short times and long times and in the limit of
zero coupling between the system and environment (isolated harmonic
oscillator). Besides the dissipation constant these expressions give the explicit
dependence on the squeezing parameter and the correlation coefficient. The
obtained uncertainty functions show explicitly the contributions of quantum and
thermal fluctuations of the system and environment. There are three
contributions to the uncertainty \cite{AndH}: (i) uncertainty which is intrinsic to
quantum mechanics, expressed through the Heisenberg uncertainty principle
(\ref{hut}),  which is not dependent on the dynamics; (ii) uncertainty that
arises due to the spreading or reassembly (the reverse of spreading) of the
wave packet, which depends on the dynamics and may increase or decrease
the uncertainty; (iii) uncertainty due to the coupling to a thermal environment,
which has two components: dissipation and diffusion (this latter is responsible
for the process of decoherence); this generally tends to increase the
uncertainty as time evolves. In the Lindblad model the uncertainty relations are
fulfilled, while in some other models considered in literature, the uncertainty
relations are violated at some initial moments of time.

We have described the evolution of the system from a quantum pure state to a
non-equilibrium quantum statistical state and to an equilibrium quantum
statistical state and we have analyzed the relaxation process. We also found the
regimes in which each type of fluctuations is important. The three stages are
marked by the decoherence time and the relaxation time, respectively. The
regime in which thermal fluctuations become comparable with the quantum
fluctuations coincides with the regime in which the decoherence effects come
into play. In other words, the system evolves from a quantum-dominated state
to a thermal-dominated state in a time which is comparable with the
decoherence time calculated in similar models in the context of transitions from
quantum to classical physics \cite {AndH,AnH,hu1,hu2}.

With this study one can understand the relation between quantum, thermal and
classical fluctuations. With the two characteristic times, namely the relaxation
time and the decoherence time to be determined in further studies on
environment-induced decoherence in the Lindblad model, one can give further
contributions in describing the role of quantum and thermal fluctuations and,
using the uncertainty relations, the transition from quantum to classical physics.
In this context we have shown recently \cite{afor,apr} that in the Lindblad
model the Schr\"odinger generalized uncertainty relation is minimized for all
times for Gaussian pure initial states of the form of correlated coherent states
for a special choice of the diffusion and dissipation coefficients. Such states are
therefore the ones that suffer the least amount of noise and they are connected
with the decoherence phenomenon \cite{paz1,paz2,zur,apr}, being the most
predictable and stable under the evolution in the presence of the environment.

Recently there is an increased interest in quantum Brownian motion as a
paradigm of quantum open systems. The possibility of preparing systems in
macroscopic quantum states leads to the problems of dissipation in tunneling
and of loss of quantum coherence (decoherence), which are intimately related
to the issue of the transition from quantum to classical physics. The Lindblad
theory provides a selfconsistent treatment of damping as a general extension of
quantum mechanics to open systems and opens more possibilities to study
these problems than the usual models of quantum Brownian motion.

{\bf Acknowledgments}

This work has partly been supported by a grant of the Romanian Ministry of
Education and Research, under contract No. A6/2001. A. I. gratefully
acknowledges this financial assistance. Financial support and hospitality at the
Institute of Theoretical Physics in Giessen during the stay of one of the authors
(A.I.) are also gratefully acknowledged.

\end{document}